\documentclass[graybox]{svmult}
\usepackage{mathptmx}       
\usepackage{helvet}         
\usepackage{courier}        
\usepackage{makeidx}         
\usepackage{graphicx}        
\usepackage{multicol}        
\usepackage[bottom]{footmisc}
\usepackage{amsmath,amssymb,amsfonts}        
\usepackage{cite}

\makeindex             

\DeclareRobustCommand{\coprod}{\mathop{\text{\fakecoprod}}}
\newcommand{\fakecoprod}{%
  \sbox0{$\prod$}%
  \smash{\raisebox{\dimexpr.9625\depth-\dp0}{\scalebox{1}[-1]{$\prod$}}}%
  \vphantom{$\prod$}%
}

\def\sqr#1#2{{\vcenter{\vbox{\hrule height.#2pt
            \hbox{\vrule width.#2pt height#1pt \kern#1pt
                  \vrule width.#2pt}\hrule height.#2pt}}}}

\def\square
 {\mathop{\mathchoice{\sqr{12}{15}}{\sqr{9}{12}}
{\sqr{6.3}{9}}{\sqr{4.5}{9}}}}

\def\sqra#1#2#3{{\vcenter{\vbox{\hrule height.#2pt
            \hbox{\vrule width.#2pt height#1pt \kern5pt 
#3
                  \vrule width.#2pt}\hrule height.#2pt}}}}

\begin{document}
\title*{An introduction to decomposition}
\author{Eric Sharpe}
\institute{Eric Sharpe \at Virginia Tech, Department of Physics MC 0435, 850 West Campus Drive, Blacksburg, VA  24061, \email{ersharpe@vt.edu} }
%
%
\maketitle

\abstract{
We review work on `decomposition,' 
a property of two-dimensional theories
with 1-form symmetries and, more generally, $d$-dimensional theories 
with $(d-1)$-form
symmetries.  Decomposition is the observation that
such quantum field theories are equivalent to (`decompose into’) 
disjoint unions of other QFTs, known in this context as ``universes.''  
Examples include two-dimensional gauge theories and
orbifolds with matter invariant under a subgroup of the gauge group.
Decomposition explains and relates several
physical properties of these theories -- for example, 
restrictions on allowed instantons arise as a ``multiverse
interference effect'' between contributions from constituent universes.  
First worked out in 2006 as part of efforts to resolve technical
questions in string propagation 
on stacks, decomposition has been the driver of a number of developments since.
We give a general overview of decomposition, describe features of decomposition
arising in gauge theories, then dive into specifics for orbifolds.  We conclude
with a discussion of the 
recent application to anomaly resolution of
Wang-Wen-Witten in two-dimensional orbifolds.
This is a contribution to the proceedings of
the conference                            
{\it Two-dimensional supersymmetric
theories and related topics} (Matrix Institute,
Australia, January 2022), giving an overview of a talk given 
there and elsewhere.
}

\section{Introduction}
Briefly, decomposition is the observation that some QFTs with a local
action are secretly equivalent to sums (disjoint unions) of other
QFTs, known in this context as `universes.'

When this happens, we say that the QFT decomposes (into its constitutent
univeres).  Decomposition of the QFT can be applied to give insight
into its properties.

Decomposition was first described in 2006 in
\cite{Hellerman:2006zs}, where it arose as part of efforts to understand
string compactifications on generalizations of spaces known as stacks
and gerbes, and resolved some of the apparent physical inconsistencies of those
theories.  It has since been developed in numerous other papers,
see for example
\cite{Robbins:2020msp,Robbins:2021lry,Robbins:2021ibx,Robbins:2021xce,
Yu:2021zmu,Tanizaki:2019rbk,Nguyen:2021yld,Nguyen:2021naa,Cherman:2020cvw,
Cherman:2021nox,Sharpe:2014tca,Sharpe:2019ddn,Sharpe:2021srf,Eager:2020rra,
Anderson:2013sia,Komargodski:2020mxz,ajt1,ajt2,ajt3,t1,gt1,xt1,
Caldararu:2010ljp,Hellerman:2010fv,Honda:2021ovk,Huang:2021zvu,
Gu:2021yek,Gu:2021beo}
and discussed in review articles including
\cite{Sharpe:2006vd,Sharpe:2010zz,Sharpe:2010iv,Sharpe:2019yag}.

For one QFT to be a sum (or disjoint union\footnote{We will use the
terms `sum' and `disjoint union' interchangeably in this article.})
of other QFTs means that, for example, there exist projection operators:
a set of topological operators $\Pi_i$ that commute with all operators
in the theory 
\begin{equation}  \label{eq:commute}
[ \Pi_i, {\cal O} ] \: = \: 0,
\end{equation}
and with the properties that
\begin{equation}  \label{eq:idempotent}
\Pi_i \Pi_j \: = \: \delta_{ij} \Pi_j, \: \: \:
\sum_i \Pi_i \: = \: 1.
\end{equation}
From~(\ref{eq:commute}), the projectors are mutually commuting,
so they can be simultaneously diagonalized:
the Fock space can be
diagonalized into eigenmodes of the projectors $\Pi_i$, which are the
states of the constituent universes.
From~(\ref{eq:idempotent}), one can show that
correlation functions are a sum of
correlation functions in the constituent theories.  Using the properties
above, we can write
\begin{eqnarray}
\langle {\cal O}_1 \cdots {\cal O}_m \rangle
& = & \sum_i \langle \Pi_i {\cal O}_1 \cdots {\cal O}_m \rangle,
\\
& = & \sum_i \langle (\Pi_i {\cal O}_1) \cdots (\Pi_i {\cal O}_m) \rangle,
\\
& = & \sum_i \langle \tilde{\cal O}_1 \cdots \tilde{\cal O}_m 
\rangle_i,
\end{eqnarray}
where the $\tilde{\cal O}$ are the projections of the operators
${\cal O}$ into the universes.

In practice, beyond exhibiting projection operators, another property
will be used frequently in this article in describing and checking 
decompositions, namely that (on a connected spacetime) the partition function
of a disjoint union is the sum of the partition functions of the
constituent universes:
\begin{equation}
Z \: = \: \sum_i Z_i.
\end{equation}
In essence, this is because the state space of the disjoint union is a sum
of the state spaces of the constituent theories. 
Formally, if we write
\begin{equation}
Z \: = \: \sum_{\rm states} \exp(-\beta H),
\end{equation}
then using the fact that the whole state (Fock) space is a sum over
state spaces of the constituent universes, we have immediately that
\begin{equation}
Z \: = \: \sum_i \sum_{\rm states \: in \: i} \exp(-\beta H)
\: = \: \sum_i Z_i.
\end{equation}

\begin{svgraybox}
Let us take a moment to distinguish universes from superselection sectors
arising in spontaneous symmetry breaking.  Briefly, the idea of a superselection
sector is that it is a sector of the theory characterized by a vacuum that 
cannot be shifted perturbatively by local operators, but can be changed
by adding energy (broadly proportional to volume).  A prototypical
example of superselection sectors is the orientation of magnetization
(microscopically, spins) in a bar
magnet.  In that example, local operators cannot change the direction of
magnetization; however, by adding energy (broadly proportional to the 
volume), one can randomize the spins, then as the magnet cools, the 
spins may align in a new direction.  Here, however, it should be noted
that the different superselection sectors (magnetizations) are linked
to one another -- except in deep IR or infinite volume limits, there is
a continuous path in field space connecting them.  By contrast,
in decomposition, one has disjoint QFTs at every
energy scale, meaning that there
is no continuous path linking states in different universes.
A more detailed discussion can be found in \cite{Tanizaki:2019rbk}.
\end{svgraybox}

Later we will work through examples of decomposition in detail,
but for the moment, let us outline some known examples.
\begin{itemize}
\item Orbifolds.  Broadly speaking, orbifolds in which a subgroup of
the gauge group acts trivially decompose, see
e.g.~\cite{Hellerman:2006zs,Robbins:2020msp,Robbins:2021lry,Robbins:2021ibx,Robbins:2021xce}.  
There is a longer story behind why orbifolds with
trivially-acting subgroups differ from ordinary orbifolds, see
\cite{Pantev:2005rh,Pantev:2005zs,Pantev:2005wj} for early work, but in
any event, much of this review will be devoted to orbifolds, so we will
see detailed examples shortly.
\item Two-dimensional gauge theories in which a subgroup of the
gauge group acts trivially also decompose.  Such gauge theories, and
their differences from ordinary gauge theories, were discussed in
\cite{Pantev:2005rh,Pantev:2005zs,Pantev:2005wj}.  Examples of their
decomposition include the following:
\begin{itemize}
\item A two-dimensional $U(1)$ gauge theory with nonminimal charges
is equivalent to a sum of $U(1)$ theories with minimal charges
\cite{Hellerman:2006zs},
\item A two-dimensional $G$ gauge theory with center-invariant matter
is equivalent to a union of $G/Z(G)$ gauge theories (for $Z(G)$ the
center of $G$) with discrete theta angles \cite{Sharpe:2014tca},
\item Two-dimensional pure Yang-Mills theory for gauge group $G$
is equivalent to a sum of invertible field theories, indexed
by irreducible representations of $G$
\cite{Nguyen:2021yld,Nguyen:2021naa} (see also \cite{Cherman:2020cvw} for the
abelian case).
\end{itemize}
\item Four-dimensional Yang-Mills theory with a restriction to instantons
of degree divisible by $k > 1$ is equivalent to a disjoint union of
$k$ ordinary four-dimensional Yang-Mills theories with different
theta angles \cite{Tanizaki:2019rbk}.  (This also, correctly, suggests
a subtlety involving cluster decomposition, to which we shall return
shortly.)
\item Unitary two-dimensional topological field theories (with semisimple
local operator algebras).  It has been known for many years that
these are equivalent to disjoint unions of invertible field theories,
see e.g.~\cite{Durhuus:1993cq,Moore:2006dw}, and by utilizing
noninvertible higher-form symmetries, it was argued in
\cite{Komargodski:2020mxz,Huang:2021zvu} 
that these are also examples of decomposition.
\item Finally, sigma models on gerbes.  Gerbes are examples of stacks,
generalizations of spaces which admit metrics, spinors, gauge fields,
and everything else one would require to make sense of a sigma model.
Sigma models with target stacks were studied in 
\cite{Pantev:2005rh,Pantev:2005zs,Pantev:2005wj}, in the hope of
discovering new string compactifications, new (2,2) SCFTs,
and amongst the technical
challenges that arose (construction of an action, presentation dependence,
naively inconsistent moduli) was, in the case of gerbes, a violation
of cluster decomposition.  The original motivation for
decomposition \cite{Hellerman:2006zs} 
was to resolve this problem.  The resolution
observed that a sigma model on a gerbe is equivalent to a disjoint union
of sigma models on spaces, solving the issue with cluster decomposition,
but also clarifying that one could not construct new (2,2) SCFTs in
this fashion.
\end{itemize}

So far we have outlined a number of rather diverse-looking examples of
decomposition, in both two and four dimensions.  The reader may well ask,
what do these examples have in common?

Briefly, in $d$ spacetime dimensions, a theory decomposes when it
has a $(d-1)$-form symmetry.  (In two dimensions, this was the
point of \cite{Hellerman:2006zs}, and it was generalized to higher
dimensions in \cite{Tanizaki:2019rbk,Cherman:2020cvw}.)  Thus,
decomposition and higher-form symmetries go hand-in-hand.

In this review, we will primarily focus on the case $d=2$, for which
one will have a decomposition if a $(d-1) = 1$-form symmetry is present.

To that end, let us take a moment to briefly review one-form symmetries.
For this review, intuitively, a one-form symmetry group is 
(something like) a group that exchanges nonperturbative sectors.

For example, consider a $G$ gauge theory or orbifold in which the
matter/fields are invariant under a subgroup $K \subset G$.
For simplicity, let us assume that $K$ is abelian, and in fact lies within
the center of $G$.  Then, in this case, there is a permutation symmetry
amongst the nonperturbative sectors.  Schematically, the path integral
is invariant under
\begin{equation}
\left( \mbox{$G$-bundle} \right) \: \mapsto \:
\left( \mbox{$G$-bundle} \right) \otimes 
\left( \mbox{$K$-bundle} \right),
\end{equation}
or if the reader prefers, in terms of gauge fields,
\begin{equation}
A \: \mapsto \: A + A',
\end{equation}
where $A$ is a $G$-instanton and $A'$ is a $K$-instanton.

Instead of an action of elements of groups, we have an action of
bundles of groups.  This structure is almost a group, except that
associativity of multiplication only holds up to multiplication.
Technically, this is known as a 2-group.

The 2-group
whose elements are $K$-bundles, is denoted either $K^{(1)}$ (recently
in physics) or $BK$ (in math).
The latter notation has been used for decades, so we will use the
notation $BK$ to denote a 2-group of $K$ bundles.

One-form symmetries can also been seen in the algebra of topological
local operators, where they are often realized nonlinearly.
This is how the decomposition story connects to two-dimensional
topological field theories \cite{Komargodski:2020mxz,Huang:2021zvu}, 
but is beyond the scope of this
article.

There are several descriptions of the two-dimensional
quantum field theories which we will
discuss in this article:
\begin{itemize}
\item A gauge theory or orbifold with a trivially-acting subgroup
(i.e.~a non-complete charge spectrum),
\item A theory with a restriction on instantons,
\item Sigma models on gerbes,
\item A theory with multiple topological local operators.
\end{itemize}

Decomposition often relates these different pictures.
\begin{itemize}
\item For one example, we will see that restrictions on instantons are
implemented as a ``multiverse interference effect'' between the different
universes of a decomposition.
\item The one-form symmetry of the quantum field theory can be
understood in the sigma model language.  A `gerbe' is a fiber bundle
whose fibers are 2-groups $BK$ of one-form symmetries.  In any case in
which the target space of a sigma model is a fiber bundle,
the sigma model possesses a global symmetry corresponding to translations
along the fibers.  For a sigma model whose target is a gerbe,
since the fibers are copies of $BK$, the sigma model has a $BK$
symmetry.
\item This is described by gauge theories with trivially-acting subgroups
because
$BK = [{\rm point}/K]$.  (In ordinary geometry, a quotient of a point by any
group is the same point back again, but the pertinent mathematics keeps track
of automorphisms, and so this is different from a point.)  Utlimately,
a sigma model whose target is a gerbe involves fibering a
trivially-acting $K$ gauge theory over an ordinary theory, hence
gauge theories with trivially-acting subgroups.
\end{itemize}

\section{Generalities on gauge theories}
Suppose we have a two-dimensional $G$ gauge theory, where $G$ is semisimple,
and a subgroup $K$ of the center of $G$ acts trivially on all the matter.

As outlined above, this theory has a global $BK$ one-form symmetry,
and so one expects that it should decompose.

The projection operators are, schematically, twist fields / Gukov-Witten
operators \cite{Gukov:2006jk,Gukov:2008sn}
corresponding to elements of the center of the group
algebra ${\mathbb C}[K]$.  Existence of projectors (idempotents), forming
a basis for the center, is a consequence of Wedderburn's theorem
(see e.g.~\cite[section XVII.3]{lang}).

In particular, judging from the projectors of Wedderburn's theorem,
universes are in one-to-one correspondence with irreducible
representations of $K$.

Now, abstractly, that is a formal argument for a decomposition, but
it does not specify the form of the decomposition.
In this particular case, decomposition takes the form
(see e.g.~\cite[section 2]{Sharpe:2014tca})
\begin{equation}
{\rm QFT}\left( G-\mbox{gauge theory}\right) \: = \:
\coprod_{\theta \in \hat{K}} {\rm QFT}\left( G/K-\mbox{gauge theory
with discrete theta angle }\theta \right).
\end{equation}

For example, $SU(2)$ gauge theory with center invariant matter is the
disjoint union of a pair of $SO(3)$ theories, schematically
\begin{equation}
SU(2) \: = \: SO(3)_+ \: + \: SO(3)_-,
\end{equation}
where the $\pm$ denotes the (${\mathbb Z}_2$-valued) discrete theta
angle coupling to the second Stiefel-Whitney class $w_2 \in H^2({\mathbb Z}_2)$.

Perturbatively, the $SU(2)$ and $SO(3)_{\pm}$ theories are identical,
but nonperturbatively, they differ.  Specifically, there are more
$SO(3)$ instantons (bundles) than $SU(2)$ instantons (bundles).
The effect of the discrete theta angle is to weight the
non-$SU(2)$ $SO(3)$ instantons by a sign, so that when the partition functions
for the $SO(3)_{\pm}$ theories are added, contributions from
non-$SU(2)$ $SO(3)$ instantons cancel out between the two theories,
leaving only $SU(2)$ instantons -- consistent with decomposition.

We can describe this more formally as follows.
Write the partition function of the disjoint union as
\begin{equation}
Z \: = \: \sum_{\theta \in \hat{K}} \int [DA] \exp(-S) \exp\left( 
\theta \int w_2(A) \right),
\end{equation}
where we use $w_2 \in H^2(K)$ to denote the degree-two characteristic class
of $G/K$ bundles.
Now, moving the summation inside the path integral, we have
\begin{eqnarray}
Z & = &
 \int [DA] \exp(-S)  \left( \sum_{\theta \in \hat{K} }
\exp\left( 
\theta \int w_2(A) \right) \right).
\end{eqnarray}
However,
\begin{equation}
 \sum_{\theta \in \hat{K} }
\exp\left( 
\theta \int w_2(A) \right)
\end{equation}
is proportional\footnote{
The proportionality factor reflects that fact that the $G$ gauge theory
has more gauge transformations than the $G/K$ gauge theory, so the
path integrals have, in principle, slightly different normalization factors.
} to a projection operator, projecting out instantons (bundles) for
which $w_2 \neq 0$, leaving only those bundles which exist in the
$G$ gauge theory.

In effect, an interference effect between the universes of 
decomposition -- a ``multiverse interference effect'' -- has cancelled out
some of the nonperturbative sectors.

As a quick consistency check, let us compare to pure $SU(2)$ Yang-Mills
theory in two dimensions, for which in essence everything is
computable \cite{Migdal:1975zg,Rusakov:1990rs,Witten:1991we}.  
We will check a decomposition due to
the $B {\mathbb Z}_2$ center symmetry \cite[section 2.4]{Sharpe:2014tca}; 
a more extreme decomposition
(to invertible field theories, utilizing noninvertible higher-form symmetries)
was discussed in \cite{Nguyen:2021yld,Nguyen:2021naa}.

Consider partition functions in pure Yang-Mills theory in two dimensions.
From \cite{Migdal:1975zg,Rusakov:1990rs,Witten:1991we},
the partition functions of the pure $SU(2)$ and the pure $SO(3)$ theory
without a discrete theta angle (denoted $SO(3)_+$), the partition functions
are of the form
\begin{eqnarray}
Z\left( G \right) & = &
\sum_R \left( \dim R \right)^{2-2g} \exp\left( - A C_2(R) \right),
\end{eqnarray}
where $G$ is the gauge group ($SU(2)$ or $SO(3)$ here),
$g$ is the genus of the two-dimensional spacetime,
$A$ its area, $C_2(R)$ the second Casimir,
and the sum is over all irreducible representations of $G$.
We also need the partition function of the
$SO(3)_-$ theory; this was computed in
\cite{Tachikawa:2013hya},
and takes the same form as above, namely
\begin{eqnarray}
Z\left( SO(3)_- \right) & = &
\sum_R \left( \dim R \right)^{2-2g} \exp\left( - A C_2(R) \right),
\end{eqnarray}
where the sum is now over representations of $SU(2)$ that are not
representations of $SO(3)$, a complementary sum to that appearing
in the $SO(3)_+$ partition function.
Assembling these pieces, we see immediately that adding a sum over
$SO(3)$ representations to a sum over $SU(2)$ representations minus
$SO(3)$ representations,
we get
\begin{equation}
Z\left( SU(2) \right) \: = \:
Z\left( SO(3)_+ \right) \: + \:
Z\left( SO(3)_- \right),
\end{equation}

\begin{svgraybox}
We have discussed a number of theories with properties such as 
a restriction on instantons, and multiple identity operators,
properties which are often taken to signal a violation of 
cluster decomposition (see e.g. \cite[section 23.6]{Weinberg:1996kr}).
Especially as cluster decomposition is sometimes used interchangeably with
locality, this is often taken to indicate a sickness or inconsistency
of the theory.

However, that is an oversimplification.
For example, cluster decomposition can also be violated in 
spontaneous symmetry breaking, in infinite-volume limits.  
This does not imply an inconsistency
of the theory, but rather is just a statement about the vacuum.
Perhaps cluster decomposition is best thought of as a property of
vacua, and locality a property of the theory, and as these examples
illustrate, these are distinct notions, not interchangeable with one
another.

In any event, the theories we are describing are manifestly local, in that they
have local Lagrangians, and the separate universes are perfectly
consistent. 
\end{svgraybox}

\section{Consistency tests and applications}
Since 2005, decomposition has been checked in a wide variety of examples,
in a wide variety of different kinds of examples, and in many ways.
We list a few examples below:
\begin{itemize}
\item Gauged linear sigma models (GLSMs):  Decomposition has been checked
in gauged linear sigma models via mirror symmetry and in quantum cohomology
rings (the latter through Coulomb branch computations).
For abelian GLSMs, this was described in \cite{Pantev:2005rh,Pantev:2005zs}
using Hori-Vafa mirrors \cite{Hori:2000kt}; 
for nonabelian GLSMs, this was checked
in the papers describing nonabelian mirror constructions
\cite{Gu:2018fpm,Chen:2018wep,Gu:2019zkw,Gu:2020ivl}.
\item In orbifolds, decomposition has been checked extensively 
\cite{Hellerman:2006zs,Robbins:2020msp,Robbins:2021lry,Robbins:2021ibx,Robbins:2021xce} in,
for example, partition functions and massless spectra, as we will outline
later in this article.
\item Decomposition has also been checked in open strings and K theory
\cite{Hellerman:2006zs}.  Briefly, in a gauge theory in which a subgroup
$K$ of the gauge group acts trivially on bulk degrees of freedom,
$K$ can still act nontrivially on boundary degrees of freedom, which 
therefore organize according to irreducible representations of $K$,
precisely matching the description of universes earlier.
In this picture, from gauge invariance, open string states can only exist
on open strings connecting the same irreducible representations of $K$ --
meaning, that there are no open strings connecting different universes.
This also can be understood in terms of K theory \cite{Witten:1998cd}.
As discussed in \cite{Hellerman:2006zs}, K theory on gerbes is equivalent
to (twisted) K theory on a disjoint union of spaces, following the same
pattern as decomposition.
\item In supersymmetric gauge theories in two dimensions, 
supersymmetric localization can
be applied to give further tests of decomposition,
as discussed in \cite{Sharpe:2014tca}.
\item In nonsupersymmetric pure Yang-Mills in two dimensions, 
decomposition can also be
checked.  We have previously outlined tests of decomposition along
center one-form symmetries, which are described in greater detail
in \cite{Sharpe:2014tca}.  In addition, there exists a more extreme
decomposition of nonsupersymmetric pure Yang-Mills to a disjoint
union of invertible field theories, indexed by irreducible representations
of the gauge group \cite{Nguyen:2021yld,Nguyen:2021naa}.
\item Decomposition in adjoint QCD$_2$ was studied in
\cite{Komargodski:2020mxz}.
\item Decomposition has been checked numerically in lattice gauge theory
\cite{Honda:2021ovk}.
\item Finally, lest we give a different impression, decomposition is not
restricted to two-dimensional theories, but has also been studied
in other dimensions, see e.g.~\cite{Tanizaki:2019rbk,Cherman:2020cvw,Cherman:2021nox}.
\end{itemize}

We should also mention that decomposition has a number of
applications:
\begin{itemize}
\item The original application was to understand and resolve certain
technical issues in making sense of sigma models whose targets are
generalized spaces known as `stacks' 
\cite{Pantev:2005rh,Pantev:2005zs,Pantev:2005wj}.  This was part of a program of
trying to construct new string compactifications, new conformal
field theories.  After one understands basic issues such as the
construction of an action to describe such a sigma model,
potential presentation-dependence issues, and moduli mismatches,
more subtle issues remain.
For example, in the 
special cases of stacks known as `gerbes' (fiber bundles whose fibers
are one-form symmetry groups), the sigma models violated cluster
decomposition.  Understanding this issue was the original motivation for
work on decomposition, and the resolution was that such sigma models
are equivalent to disjoint unions of sigma models on ordinary spaces.
As a result, we were not able to construct new (2,2) supersymmetric SCFTs,
though we did learn about decomposition.  (In the more general
case of (0,2) SCFTs for gerbes, it is still an open question of whether new
string compactifications exist, see e.g.~\cite{Anderson:2013sia}.)
\item Decomposition makes predictions for Gromov-Witten theory,
specifically the Gromov-Witten theory of stacks and gerbes
\cite{Chen:2000cy,Chen:2000kya,Abramovich:2001vh}.
From decomposition, the Gromov-Witten theory of a gerbe must match
that of a disjoint union of spaces, and this was checked rigorously 
and discussed in
e.g.~\cite{ajt1,ajt2,ajt3,t1,gt1,xt1}.
\item In gauged linear sigma models, decomposition was used to
provide a novel nonperturbative construction of branched double covers
in \cite{Caldararu:2010ljp}, giving examples of GLSMs with nonbirational
phases, realizing examples of Kuznetsov's
homological projective duality \cite{kuz2}.  This construction has been
utilized in the GLSM community in a number of examples since,
see e.g.~\cite{Hori:2011pd,Chen:2020iyo,Guo:2021aqj} for a few
examples.
\item Applications to computing elliptic genera of pure gauge theory,
studying IR limits of pure supersymmetric gauge theories in two dimensions
\cite{Aharony:2016jki}, were discussed in \cite{Eager:2020rra}.
\item Recently decomposition has been applied 
\cite{Robbins:2021lry,Robbins:2021ibx,Robbins:2021xce} to understand and
simplify the Wang-Wen-Witten anomaly resolution proposal
\cite{Wang:2017loc}, as we shall discuss in section~\ref{sect:www}.
\end{itemize}

\section{Multiverse interference, portals, and wormholes}
Next, let us summarize some of the more entertaining features of
decomposition:
\begin{itemize}
\item {\it Multiverse interference effects.}  
We have already seen that summing over
universes has the effect of projecting out some nonperturbative contribution,
hence a ``multiverse interference effect.'' Our primary exmaples was of
a two-dimensional $SU(2)$ gauge theory with center-invariant matter, for which,
schematically,
\begin{equation}
{\rm QFT}\left( SU(2) \right) \: = \:
{\rm QFT}\left( SO(3)_+ \right) \, \coprod \,
{\rm QFT}\left( SO(3)_- \right).
\end{equation}
\item {\it Fundamentally-charged Wilson lines are defects bridging universes.}
Consider for example two-dimensional abelian $BF$ theory at level $k$.
This theory decomposes into a disjoint union of $k$ invertible field theories.
The projectors are
\begin{equation}
\Pi_m \: = \: \frac{1}{k} \sum_{n=0}^{k-1} \xi^{nm} {\cal O}_n,
\end{equation}
where $\xi = \exp(2 \pi i/k)$ and
\begin{equation}
{\cal O}_n \: = \: : \exp\left( n B \right) :.
\end{equation}
These local operators have clock-shift commutation relations
with the Wilson lines (see e.g.~\cite{Hellerman:2010fv})
\begin{equation}
{\cal O}_p W_q \: = \: \xi^{pq} W_q {\cal O}_p,
\end{equation}
which algebraically are equivalent to
\begin{equation}
\Pi_m W_p \: = \: W_p \Pi_{m + p \mod k}.
\end{equation}
Thus, moving a projector past a Wilson line changes the projector,
and so Wilson lines in abelian $BF$ theory act as (nondynamical) defects
bridging different universes.
\item {\it Wormholes between universes.}  This is how GLSMs realize branched
double covers nonperturbatively \cite{Caldararu:2010ljp}.  
Consider for example a GLSM with
gauge group $U(1)$, two chiral superfields $p_a$ of charge $+2$ and
four chiral superfields $\phi_i$ of charge $-1$, with a superpotential
\begin{equation}
W \: = \: \sum_{ij} \phi_i \phi_j A^{ij}(p).
\end{equation}
In the phase $r \ll 0$, this describes a branched double cover of
${\mathbb P}^1$, where the sheets of the cover are (approximate) universes,
spanned by the $p$ fields (of nonminimal charge), and the branch locus
is the region where the mass matrix $A^{ij}$ develops zero eigenvalues,
forming a Euclidean wormhole.
\end{itemize}

\section{Specifics on orbifolds}
Now, let us turn to examples of decomposition in two-dimensional orbifolds.

Consider an orbifold $[X/\Gamma]$, where $K \subset \Gamma$ acts trivially,
and let $G = \Gamma/K$.
We can write $\Gamma$ as
\begin{equation}
1 \: \longrightarrow \: K \: \longrightarrow \: \Gamma \: \longrightarrow \:
G \: \longrightarrow \: 1.
\end{equation}
Decomposition is known for general extensions \cite{Hellerman:2006zs},
but for simplicity in this overview, let us assume for the moment
that this is a central extension, so that $K$ is a subset of the
center of $\Gamma$.  Let $[\omega] \in H^2(G,K)$ denote the
element of group cohomology classifying the extension.

In this case, if $K$ acts trivially on $X$ and lies within the
center of $\Gamma$, then this orbifold decomposes as
\begin{equation}  \label{eq:decomp-orb}
{\rm QFT}\left( [X/\Gamma] \right) \: = \:
{\rm QFT}\left( \coprod_{\rho \in \hat{K}} 
[X/G]_{\hat{\omega}(\rho)} \right),
\end{equation}
where $\hat{K}$ denotes the set of isomorphism classes of irreducible
representations of $K$, and $\hat{\omega}(\rho)$ represent
discrete torsion phases, essentially finite-group analogues of
theta angles, which is the image of the extension class under $\rho$:
\begin{eqnarray}
H^2(G,K) & \stackrel{\rho}{\longrightarrow} & H^2(G,U(1)), 
\nonumber \\
\omega & \mapsto & \omega \circ \rho \: = \: \hat{\omega}(\rho).
\nonumber
\end{eqnarray}

This will be a finite-group analogue of the $SU(2)$ decomposition 
described earlier, with $[X/\Gamma]$ playing the analogue of the
$SU(2)$ theory and the $[X/G]_{\hat{\omega}(\rho)}$ theories
playing the analogue of the $SO(3)_{\pm}$ theories.

To justify this decomposition, we must first provide projectors.
Corresponding to any irreducible representation $R \in \hat{K}$,
the projector is \cite{Sharpe:2021srf}
\begin{equation}  \label{eq:genl-projectors}
\Pi_R \: = \: \sum_i \frac{ \dim R_i }{|K|} \sum_{k \in K} 
\chi_{R_i}\left( k^{-1} \right) \tau_k,
\end{equation}
where $\tau_k$ is a twist field for the trivially-acting element
$k \in K$, $\chi_R(g)$ denotes the character of $g$ in representation $R$,
and the $R_i$ are a set of representatives of the irreducible representations.
(For a detailed examination of why trivially-acting group elements
have associated twist fields, and the unitarity violations that ensue if
one assumes otherwise, see \cite{Pantev:2005rh}.)
It can be shown that these projectors have the expected properties, namely
\begin{equation}
\Pi_R \Pi_S \: = \: \delta_{R,S} \Pi_R, \: \: \:
\sum_R \Pi_R \: = \: 1.
\end{equation}

As the twist fields in this case can be understood formally as the
center of the group algebra, this expression for the projector is a formal
consequence of Wedderburn's theorem in mathematics 
(see e.g.~\cite[section XVII.3]{lang}).

To make this more concrete, let us examine all the details in one
particular example (taken from
\cite[section 5.2]{Hellerman:2006zs}).  
Take $\Gamma = D_4$, the eight-element dihedral
group, with center $K = {\mathbb Z}_2$, which we will assume
acts trivially on $X$.
In this case, decomposition~(\ref{eq:decomp-orb}) predicts
\begin{equation} \label{eq:decomp:d4-ex}
{\rm QFT}\left( [X/D_4] \right) \: = \: 
{\rm QFT}\left( [X/{\mathbb Z}_2 \times {\mathbb Z}_2]_{\rm w/o \: d.t.}
\right) \, \coprod \,
{\rm QFT}\left( [X/{\mathbb Z}_2 \times {\mathbb Z}_2]_{\rm d.t.} \right),
\end{equation}
a disjoint union of two ${\mathbb Z}_2 \times {\mathbb Z}_2$ orbifolds,
one with discrete torsion, and the other without.

In passing, note that this is a very precise analogue of the earlier
example of an $SU(2)$ gauge theory (compare $[X/D_4]$) decomposing into
a pair of $SO(3)$ theories (compare $[X/{\mathbb Z}_2 \times {\mathbb Z}_2]$).

Now, let us check this statement.
First, we consider projection operators.  If let let $z \in D_4$ denote
the generator of the (trivially-acting) ${\mathbb Z}_2$ center,
and $\hat{z}$ the corresponding twist field, then $\hat{z}^2 = 1$,
and the projectors are
\begin{equation}
\Pi_{\pm} \: = \: \frac{1}{2} \left( 1 \pm \hat{z} \right),
\end{equation}
which obey 
\begin{equation}
\Pi_{\pm}^2 \: = \: \Pi_{\pm}, \: \: \:
\Pi_{\pm} \Pi_{\mp} \: = \: 0, \: \: \:
\Pi_+ + \Pi_- \: = \: 1.
\end{equation}
These projectors are in principle the specialization 
of~(\ref{eq:genl-projectors}) to this
case, but in fact here are sufficiently simple that they can be seen
by inspection,
as indeed was the case in \cite{Hellerman:2006zs}.

So far we have produced a pair of local projection operators, which tell
us that the theory breaks into two pieces, but to 
verify the decomposition~\ref{eq:decomp:d4-ex}, we need more information
about the pieces.  To that end, we will next compute the partition
function of this orbifold.

To compute the partition function, we need to describe the dihedral
group more explicitly.  Briefly, it is generated by elements $z$ (generating
the center), $a$, and $b$, whose products we list below:
\begin{equation}
D_4 \: = \: \{1, z, a, b, az, bz, ab, ba = abz \}.
\end{equation}
Let us compute the partition function on $T^2$.
For any orbifold $[X/\Gamma]$, the partition function on $T^2$ is
\begin{equation}
Z_{T^2}\left( [X/\Gamma] \right) \: = \:
\frac{1}{|\Gamma|} \sum_{gh = hg} Z_{g,h},
\end{equation}
where each $Z_{g,h}$ represents the path integral contribution from
the $\Gamma$ nonperturbative sector (``twisted sector'')
defined by the commuting
pair $g, h \in G$.  (Since $\Gamma$ is finite, there is no perturbative
contribution to the $\Gamma$ gauge theory, only nonperturbative
contributions.)  Schematically, each $Z_{g,h}$ is a path integral sum
over maps $T^2 \rightarrow X$ with branch cuts defined by $g, h$:
\begin{equation}
Z_{g,h} \: = \: \left(
{\scriptstyle g} \square_h \: \longrightarrow \: X 
\right)
\end{equation}
(In order for the corners of the square to close, one only sums over
commuting $g, h \in G$.)

We will argue that
\begin{equation}
Z_{T^2}\left( [X/D_4] \right) \: = \:
Z_{T^2}\left( [X/{\mathbb Z}_2 \times {\mathbb Z}_2] \right)
\: + \:
Z_{T^2}\left( [X/{\mathbb Z}_2 \times {\mathbb Z}_2]_{\rm d.t.} \right)
\end{equation}
verifying decomposition~(\ref{eq:decomp:d4-ex}).

To that end, first note that since $z$ acts trivially and
each $Z_{g,h}$ only depends upon boundary conditions, we immediately have
that
\begin{equation}
Z_{g,h} \: = \: {\scriptstyle g} \square_h \: = \:
{\scriptstyle gz} \square_h \: = \: 
{\scriptstyle g} \square_{hz} \: = \:
{\scriptstyle gz} \square_{hz} \: = \: Z_{gz,hz}.
\end{equation}
Each square ${\scriptstyle g} \square_h$ can be associated
(modulo automorphisms) with a $\Gamma$ bundle, so this is a symmetry amongst
the nonperturbative sectors of the $\Gamma$ orbifold.
Furthermore, those sectors are related by tensoring in a $B {\mathbb Z}_2$
bundle:
\begin{equation}
Z_{g,h} \: = \:
{\scriptstyle g} \square_h 
\: \: \stackrel{ {\scriptstyle z} \square_1 }{\longrightarrow}
\: \:
{\scriptstyle gz} \square_h \: \:
\stackrel{ {\scriptstyle z} \square_z }{\longrightarrow}
\: \:
{\scriptstyle g} \square_{hz} \: \:
\stackrel{ {\scriptstyle z} \square_1 }{\longrightarrow}
\: \:
{\scriptstyle gz} \square_{hz}
\: = \: Z_{gz,hz}
\end{equation}

\begin{svgraybox}
This is the $B {\mathbb Z}_2$ one-form symmetry, explicitly.
\end{svgraybox}

Next, to help clarify, write the elements of
${\mathbb Z}_2 \times {\mathbb Z}_2 = D_4/{\mathbb Z}_2$ as
\begin{equation}
{\mathbb Z}_2 \times {\mathbb Z}_2 \: = \: \{1,
\overline{a}, \overline{b}, \overline{a} \overline{b} \},
\end{equation}
where $\overline{a}$ is the projection of $\{a, az\}$,
and $\overline{b}$ is the projection of $\{b, bz\}$.
Then, we see that each $D_4$ twisted sector ($Z_{g,h}$) that appears is
the same as a $D_4/{\mathbb Z}_2 = {\mathbb Z}_2 \times {\mathbb Z}_2$
twisted sector, {\it except} for the sectors
\begin{equation}  \label{eq:d4:omitted}
{\scriptstyle \overline{a}} \square_{\overline{b}},
\: \: \:
{\scriptstyle \overline{a}} \square_{\overline{a} \overline{b}},
\: \: \:
{\scriptstyle \overline{b}} \square_{\overline{a} \overline{b}},
\end{equation}
which do not appear, because their lifts do not commute in $D_4$.
(These form a modular orbit -- modular invariance is ensured at every step.)

\begin{svgraybox}
This is a restriction on the nonperturbative sectors.
\end{svgraybox}

So far, we have argued that
\begin{eqnarray}
Z_{T^2}\left( [X/D_4] \right)
& = &
\frac{ | {\mathbb Z}_2 \times {\mathbb Z}_2 | }{ |D_4| }
| {\mathbb Z}_2 |^2
\left( Z_{T^2}\left( [X/{\mathbb Z}_2 \times {\mathbb Z}_2 ] \right)
\: - \: \left( \mbox{some twisted sectors} \right) \right),
\nonumber \\
& = &
2 \left( Z_{T^2}\left( [X/{\mathbb Z}_2 \times {\mathbb Z}_2 ] \right)
\: - \: \left( \mbox{some twisted sectors} \right) \right).
\label{eq:d4-part:p1}
\end{eqnarray}

\begin{svgraybox}
In particular, despite the fact that the ${\mathbb Z}_2$ acts trivially,
this is a different theory than the ${\mathbb Z}_2 \times {\mathbb Z}_2$
orbifold.  {\it Physics knows when we gauge even a trivially-acting group.}
\end{svgraybox}

We can simplify the expression above via the use of discrete torsion
\cite{Vafa:1986wx},
which is a set of modular-invariant phases that one can add to partition
functions -- in essence, theta angles for the finite gauge theory.
The new partition function in a $G$ orbifold on $T^2$, for example,
has the form
\begin{equation}
Z_{T^2}\left( [X/G] \right) \: = \:
\frac{1}{|G|} \sum_{gh = hg} \epsilon(g,h) Z_{g,h},
\end{equation}
where $\epsilon(g,h)$ represent the discrete torsion phases.

Now, in a $G$ orbifold, possible choices of discrete torsion phases
are classified by $H^2(G,U(1))$.  In the case $G = {\mathbb Z}_2 \times
{\mathbb Z}_2$,
\begin{equation}
H^2( {\mathbb Z}_2 \times {\mathbb Z}_2, U(1)) \: = \:
{\mathbb Z}_2,
\end{equation}
and the twisted sectors that get a phase (specifically, a sign)
are
\begin{equation}
{\scriptstyle \overline{a}} \square_{\overline{b}},
\: \: \: 
{\scriptstyle \overline{a}} \square_{\overline{a} \overline{b}},
\: \: \:
{\scriptstyle \overline{b}} \square_{\overline{a} \overline{b}},
\end{equation}
the same sectors that are omitted from the ${\mathbb Z}_2 \times {\mathbb Z}_2$
orbifold partition functino in the description of the
$D_4$ orbifold partition function in~(\ref{eq:d4-part:p1}).
Thus, we see that the $D_4$ partition function on $T^2$ can be rewritten as
\begin{equation}
Z_{T^2}\left( [X/D_4] \right) \: = \:
Z_{T^2}\left( [X/{\mathbb Z}_2 \times {\mathbb Z}_2] \right)
\: + \:
Z_{T^2}\left( [X/{\mathbb Z}_2 \times {\mathbb Z}_2]_{\rm d.t.} \right).
\end{equation}
This matches the prediction of decomposition~(\ref{eq:decomp:d4-ex}) 
in this case.

\begin{svgraybox}
In particular, adding the universes has the effect of cancelling out
some of the nonperturbative sectors, namely those listed 
in~(\ref{eq:d4:omitted}).
This is an example of a {\it multiverse interference effect.}
\end{svgraybox}

So far, we have verified that partition functions on $T^2$ reproduce
decomposition.  Analogous computations on higher-genus Riemann surfaces
also reproduce decomposition, though the combinatorics is more complex.
See \cite[section 5.2]{Hellerman:2006zs} for details.

Now, let us turn to massless spectrum computations.
(In principle, this is all implicit in the partition function computations,
but we find it instructive to explicitly study this particular facet.)
For the case $X = T^6$, with a standard ${\mathbb Z}_2 \times
{\mathbb Z}_2$ action \cite{Vafa:1994rv},
the massless spectrum of $[T^6/D_4]$ is easily computed and given by
\begin{equation}
\begin{array}{ccccccc}
 & & & 2 & & & \\
 & & 0 & & 0 & &  \\
 & 0 & & 54 & & 0 & \\
2 & & 54 & & 54 & & 2 \\
 & 0 & & 54 & & 0 & \\
 & & 0 & & 0 & & \\
 & & & 2 & & &
\end{array}
\end{equation}

This result is problematic -- the $2$'s in the corners signal a violation of
cluster decomposition, which ordinarily would be reason to believe that the
result is incorrect.  However, as we have seen, cluster decomposition arises
in any theory describing a disjoint union of QFTs, and so this is not
unexpected.  Hand-in-hand, the $2$'s can also be interpreted to mean that the
theory has two components.  In particular, the massless spectra of
the $[T^6 / {\mathbb Z}_2 \times {\mathbb Z}_2 ]$ orbifolds, with and
without discrete torsion, are given by  \cite{Vafa:1994rv}
\begin{equation}
\begin{array}{ccccccc}
 & & & 1 & & & \\
 & & 0 & & 0 & & \\
 & 0 & & 51 & & 0 & \\
1 & & 3 & & 3 & & 1 \\
 & 0 & & 51 & & 0 & \\
 & & 0 & & 0 & & \\
 & & & 1 & & &
\end{array}
\: \: \: + \: \: \:
\begin{array}{ccccccc}
 & & & 1 & & & \\
 & & 0 & & 0 & & \\
 & 0 & & 3 & & 0 & \\
1 & & 51 & & 51 & & 1 \\
 & 0 & & 3 & & 0 & \\
 & & 0 & & 0 & & \\
 & & & 1 & & &
\end{array}
\end{equation}
It is easy to see that the sum of the two $[T^6/{\mathbb Z}_2 \times
{\mathbb Z}_2]$ orbifold spectra matches that of $[T^6/D_4]$,
verifying decomposition~(\ref{eq:decomp:d4-ex}).

This example was not a one-off, but in fact verifies the general 
prediction of \cite{Hellerman:2006zs} for orbifolds with trivially-acting
subgroups.

In most of this review we have focused on cases of gauge theories and
orbifolds in which the trivially-acting subgroup is in the center,
but more general examples exist and have been studied.
Decomposition in the more general case is the statement
\cite{Hellerman:2006zs}
\begin{equation}
{\rm QFT}\left( [X/\Gamma] \right) \: = \: 
{\rm QFT}\left( \left[ \frac{ X \times \hat{K} }{G} \right]_{\hat{\omega}}
\right),
\end{equation}
where the discrete torsion on universes $\hat{\omega}$ is described
in \cite{Hellerman:2006zs}.
In the general case, where the trivially-acting $K \subset \Gamma$
need not be central, $G = \Gamma/K$ can act nontrivially on the set of
isomorphism classes of irreducible representations $\hat{K}$, and the
universes are identified with orbits of $G$ in $\hat{K}$.  In the special
case of central extensions, the $G$ action on $\hat{K}$ is trivial,
the orbits are single elements of $\hat{K}$, and the decomposition reduces
to a disjoint union of copies of $[X/G]_{\hat{\omega}}$, indexed
by $\hat{K}$, as described earlier in~(\ref{eq:decomp-orb}).

For example, consider the orbifold $[X/{\mathbb H}]$,
where ${\mathbb H}$ is the eight-element group of unit quaternions,
and where $K = \langle i \rangle \subset {\mathbb H}$ acts
trivially.  In this case, the center of ${\mathbb H}$ is
${\mathbb Z}_2$, which is contained within $K \cong {\mathbb Z}_4$,
but $K$ also has elements that are not central.
On the set $\hat{K}$, $G$ leaves two elements invariant but exchanges
two elements, so that there are a total of three $G$ orbits, and
three universes.

As is discussed in detail in \cite[section 5.4]{Hellerman:2006zs},
in this case decomposition predicts
\begin{equation}
{\rm QFT}\left( [X/{\mathbb H} \right)
\: = \: 
{\rm QFT}\left( X \, \coprod [X/{\mathbb Z}_2] \, \coprod \,
[X/{\mathbb Z}_2] \right),
\end{equation}
which is checked by exhibiting projection operators, computing
partition functions at arbitrary genus, and comparing massless spectra.
In particular, the universes in this example are not all just the same
orbifold with different choices of discrete torsion, but rather they
are qualitatively different from one another.

So far we have outlined decomposition in ordinary orbifolds
with trivially-acting subgroups.  Next we outline decomposition
in such orbifolds in the presence of discrete torsion,
following \cite{Robbins:2020msp}.

Consider an orbifold $[X/\Gamma]_{\omega}$,
where $K \subset \Gamma$ acts trivially, $\omega \in H^2(\Gamma,U(1))$
is an element of discrete torsion, and we define $G = \Gamma/K$,
so that
\begin{equation}
1 \: \longrightarrow \: K \: \stackrel{\iota}{\longrightarrow} \:
\Gamma \: \stackrel{\pi}{\longrightarrow} \: G \:
\longrightarrow \: 1.
\end{equation}
For simplicity, we assume that this is a central extension
(that $K$ maps to a subgroup of the center of $\Gamma$).
It will be helpful to utilize the maps below:
\begin{equation} \label{eq:five-term}
H^2(G,U(1)) \: \stackrel{\pi^*}{\longrightarrow} \:
\left( {\rm Ker}\, \iota^* \subset H^2(\Gamma,U(1)) \right)
\: \stackrel{\beta}{\longrightarrow} \:
H^1(G,H^1(K,U(1))) \: = \: {\rm Hom}(G,\hat{K}).
\end{equation}
Then, we can describe decomposition of $[X/\Gamma]_{\omega}$ in terms
of the following cases:
\begin{enumerate}
\item If $\iota^* \omega \neq 0$, then
\begin{equation}
{\rm QFT}\left( [X/\Gamma]_{\omega} \right)
\: = \: {\rm QFT}\left( \left[ \frac{ X \times \hat{K}_{\iota^* \omega}}{G}
\right]_{\hat{\omega}} \right).
\end{equation}
\item If $\iota^* \omega = 0$ and $\beta(\omega) \neq 0$, then
\begin{equation}  \label{eq:decomp:dt:case2}
{\rm QFT}\left( [X/\Gamma]_{\omega} \right)
\: = \: {\rm QFT}\left(
\left[ \frac{ X \times \widehat{ {\rm Coker}\ \beta(\omega) } }{ {\rm Ker}\,
\beta(\omega) } \right]_{\hat{\omega}} \right).
\end{equation}
\item If $\iota^* \omega = 0$ and $\beta(\omega) = 0$,
then $\omega = \pi^* \overline{\omega}$ for some
$\overline{\omega} \in H^2(G,U(1))$ and
\begin{equation}
{\rm QFT}\left( [X/\Gamma]_{\omega} \right)
\: = \: {\rm QFT}\left(
\left[ \frac{ X \times \hat{K} }{G} \right]_{\hat{\omega} + 
\overline{\omega}} \right),
\end{equation}
essentially the same decomposition as in the case of no discrete torsion,
but with the discrete torsion on the universes shifted by
$\overline{\omega}$.
\end{enumerate}

This description was developed more systematically in
\cite{Robbins:2020msp}, which also checked the results in numerous
examples.

\section{Quantum symmetries in noneffective orbifolds}

As one final prerequisite before describing the Wang-Wen-Witten
anomaly resolution procedure in orbifolds \cite{Wang:2017loc},
we describe some novel modular-invariant phases that one can add
to noneffective orbifolds, which were deemed
`quantum symmetries' in \cite{Robbins:2021ibx} 
(see also \cite{Tachikawa:2017gyf}) as they generalize 
the notion of quantum symmetries in orbifolds from the late 1980s.

Briefly, these phases describe actions on twisted sector states.
Consider an orbifold $[X/\Gamma]$ in which a subgroup $K \subset \Gamma$
acts trivially, and define $G = \Gamma/K$, as before.
For simplicity, assume that $K$ is central in $\Gamma$, so that
\begin{equation}
1 \: \longrightarrow \: K \: \longrightarrow \: \Gamma \: \longrightarrow \:
G \: \longrightarrow \: 1
\end{equation}
is a central extension.
In this case, the quantum symmetries (as the term is used 
in \cite{Robbins:2021ibx})
are classified by 
$H^1(G,H^1(K,U(1))) = {\rm Hom}(G,\hat{K})$, and describe an action of
$K$ on $G$-twisted sector states.  Schematically, in terms of path
integral data,
\begin{equation}
{\scriptstyle gz} \square_h \: = \: B( \pi(h), z) 
\left( {\scriptstyle g} \square_h \right),
\end{equation}
for $B \in H^1(G,H^1(K,U(1)))$, $z \in K$, and $g, h \in \Gamma$.

It will be useful to note that the gruop classifying quantum symmetries
fits into an exact sequence \cite{hochschild77}
\begin{equation} \label{eq:seven-term}
\left( {\rm Ker}\, \iota^* \subset H^2(\Gamma,U(1)) \right)
\: \stackrel{\beta}{\longrightarrow} \:
H^1(G,H^1(K,U(1))) \: \stackrel{d_2}{\longrightarrow} \:
H^3(G,U(1)).
\end{equation}
(Technically, this is part of a seven-term sequence slightly extending the
inflation-restriction sequence.)  The map $d_2$ is a differential in the 
Lyndon-Hochschild-Serre spectral sequence,
and for any $\omega \in {\rm Ker}\, \iota^*$,
\begin{equation}
\beta(\omega)(\pi(g),z) \: = \: \frac{\omega(g,z) }{ \omega(z,g) },
\end{equation}
for $g \in \Gamma$, $z \in K$.
(This is the same map $\beta$ that appeared in~(\ref{eq:five-term}).)

The first term in the sequence above can be interpreted in terms of
discrete torsion, and as in two-dimensional orbifolds $[X/G]$ gauge anomalies
are counted by $H^3(G,U(1))$, the last term can be interpreted
in terms of anomalies, so the sequence~(\ref{eq:seven-term}) can be
represented schematically as
\begin{equation}
\left( \mbox{discrete torsion} \right) \: \stackrel{\beta}{\longrightarrow} \:
\left( \mbox{quantum symmetries} \right) \: \stackrel{d_2}{\longrightarrow} \:
\left( \mbox{anomalies} \right).
\end{equation}

As this sequence suggests, those quantum symmetries in the image of $\beta$
are equivalent to choices of discrete torsion, and in fact are equivalent
to quantum symmetries in the older sense of the term.

For applications to Wang-Wen-Witten, we will need quantum symmetries
$B$ such that $d_2(B) \neq 0$.  As the sequence above is exact, such
quantum symmetries are necessarily not equivalent to discrete torsion.

Before going on, let us briefly describe decomposition in orbifolds
with quantum symmetries.  Briefly,
\begin{equation} \label{eq:decomp:qs}
{\rm QFT}\left( [X/\Gamma]_B \right) \: = \:
{\rm QFT}\left( \coprod_{ \widehat{ {\rm Coker}\, B} } [ X / {\rm Ker}\, B 
]_{\hat{\omega}} \right),
\end{equation}
where we interpret the quantum symmetry $B$ as an element of
Hom$(G,\hat{K})$.
Decomposition in this case is more or less uniquely dictated by results
for decomposition in orbifolds with discrete torsion such that
$\iota^* \omega = 0$ and $\beta(\omega) \neq 0$,
as described earlier in~(\ref{eq:decomp:dt:case2}).  It was also
checked in numerous examples in \cite{Robbins:2021ibx}.

\section{Wang-Wen-Witten anomaly resolution procedure}
\label{sect:www}
In \cite{Wang:2017loc},
Wang, Wen, and Witten proposed an algorithm to resolve gauge anomalies
in anomalous orbifolds $[X/G]$.  In this section we will review that
procedure, and then observe how decomposition gives an alternative
interpretation of the result that clarifies why the procedure removes
the anomaly.

We will use the fact that
in two dimensions, gauge anomalies in finite $G$ gauge theories
are classified by elements of $H^3(G,U(1))$.  In the anomalous
orbifold $[X/G]$, we will let $\alpha \in H^3(G,U(1))$ denote the
anomaly.

The Wang-Wen-Witten procedure has two steps:
\begin{enumerate}
\item We replace $G$ by a larger group $\Gamma$,
\begin{equation}
1 \: \longrightarrow \: K \: \longrightarrow \: \Gamma \: 
\stackrel{\pi}{\longrightarrow} \: G \: \longrightarrow \: 1,
\end{equation}
which we will assume is a central extension, and where $K$ acts
trivially.

From decomposition, if all we did was to replace $G$ by $\Gamma$,
we would not have resolved the orbifold, as physically
$[X/\Gamma]$ is equivalent to copies and covers of $[X/G]$, as we have
seen previously.
\item The second step of the Wang-Wen-Witten procedure is to turn on
a quantum symmetry phase $B \in H^1(G, H^1(K,U(1)))$, chosen so that
$d_2 B = \alpha$.  This implies that $\pi^* \alpha \in H^3(\Gamma,U(1))$
is trivial.  
\end{enumerate}
These two choices together -- an extension $\Gamma$ plus a choice
of suitable quantum symmetry $B$ -- resolve the anomaly.

From decomposition, we can see how the anomaly is resolved.
Recall from~(\ref{eq:decomp:qs} that
\begin{equation}
{\rm QFT}\left( [X/\Gamma]_B \right) \: = \:
{\rm QFT}\left( \coprod_{ \widehat{ {\rm Coker}\, B} } [ X / {\rm Ker}\, B 
]_{\hat{\omega}} \right).
\end{equation}
As we chose the quantum symmetry $B$ so that $d_2 B = \alpha$,
we have immediately that
\begin{equation}
\left. \alpha \right|_{{\rm Ker}\, B} \: = \: 0.
\end{equation}
Thus, each orbifold $[X/{\rm Ker}\, B]$ is automatically anomaly-free.

Put another way, the result of the Wang-Wen-Witten procedure -- replacing
$[X/G]$ by a larger orbifold $[X/\Gamma]_B$ -- is equivalen to replacing
$[X/G]$ by a collection of smaller orbifolds $[X/{\rm Ker}\, B]$,
in which Ker $B \subset G$ is non-anomalous.

Let us see this explicitly in examples.
We will consider several resolutions of orbifolds of the form
$[X/G]$ for $G = {\mathbb Z}_2 \times {\mathbb Z}_2$.
For this group, $H^3(G,U(1)) = ( {\mathbb Z}_2 )^3$,
corresponding to the three ${\mathbb Z}_2$ subgroups,
so if we write 
\begin{equation}
G \: = \: {\mathbb Z}_2 \times {\mathbb Z}_2 \: = \:
\{1, a, b, ab \},
\end{equation}
then
\begin{equation}
H^3(G,U(1)) \: = \: ( {\mathbb Z}_2)^3 \: = \:
\langle a \rangle \, \times \, \langle b \rangle \, \times \,
\langle a b \rangle.
\end{equation}
To apply the Wang-Wen-Witten procedure, one must make two choices,
\begin{itemize}
\item a choice of larger gauge group $\Gamma$, and
\item a choice of quantum symmetry.
\end{itemize}
We will list several examples of larger group $\Gamma$, and for each
$\Gamma$, all possible choices of quantum symmetry and the resulting
theories.

For our first resolution, we take $\Gamma = D_4$,
\begin{equation}
1 \: \longrightarrow \: {\mathbb Z}_2 \: \longrightarrow \: D_4 \:
\longrightarrow \: {\mathbb Z}_2 \times {\mathbb Z}_2 \:
\longrightarrow \: 1.
\end{equation}
The quantum symmetry $B$ is determined by its image on the generators
$\{a, b\}$, and we list all possibilites in table~\ref{table:d4}
(including cases in which we turn on discrete torsion in the $[X/\Gamma]$
orbifold, in addition to the quantum symmetry).

\begin{table}[h]
\begin{center}
\caption{\label{table:d4}
A list of all possible quantum symmetries, anomalies resolved, and
corresponding orbifolds $[X/\Gamma]_B$ for the case $\Gamma = D_4$.
The first two columsn describe the quantum symmetry; the column $d_2(B)$
gives the image of the quantum symmetry, and hence the anomaly that
can be resolved; and the last two columns give the physical theory
equivalent to $[X/\Gamma]_B$, for either choice of discrete torsion
in the $\Gamma = D_4$ orbifold.
}
\begin{tabular}{cc|c|c|c}
\hline
$B(a)$ & $B(b)$ & $d_2(B)$ & $[X/\Gamma]_B$ w/o d.t. & $[X/\Gamma]_B$
with d.t. \\ \hline
1 & 1 & $-$ & $[X/G] \coprod [X/G]_{\rm d.t.}$ & $[X/\langle b \rangle]$
\\
-1 & 1 & $-$ & $[X/\langle b \rangle]$ &
$[X/G] \coprod [X/G]_{\rm d.t.}$ 
\\
1 & -1 & $\langle b \rangle$ & $[X/\langle b \rangle]$ &
$[X/\langle ab \rangle]$ 
\\
-1 & -1 & $\langle b \rangle$ & $[X/\langle ab \rangle]$ &
$[X/\langle a \rangle]$
\end{tabular}
\end{center}
\end{table}

The first row of table~\ref{table:d4} describes the case of 
no quantum symmetry.  Nothing is 
resolved, and the physical theories are (copies of) the $G$ orbifold.
The last two rows are more interesting.  These describe cases in which
an anomaly in the subgroup $\langle b \rangle \subset G$ can be resolved.
The resulting physical theories, listed in the last two columns (corresponding
to either choice of discrete torsion in the $\Gamma = D_4$ orbifold) are
orbifolds by subgroups not containing $\langle b \rangle$, and so by
assumption are anomaly-free.  In particular, we see that the Wang-Wen-Witten
prescription works, explicitly.

For our next resolution of the anomalous $[X/G]$ orbifold,
for the same $G$ as before, we take $\Gamma = {\mathbb H}$, the
eight-element group of unit quaternions,
\begin{equation}
1 \: \longrightarrow \: {\mathbb Z}_2 \: \longrightarrow \:
{\mathbb H} \: \longrightarrow \: {\mathbb Z}_2 \times {\mathbb Z}_2
\: \longrightarrow \: 1.
\end{equation}
We list all possibilites for the quantum symmetry in table~\ref{table:h}.

\begin{table}[h]
\begin{center}
\caption{\label{table:h}
A list of all possible quantum symmetries, anomalies resolved, and
corresponding orbifolds $[X/\Gamma]_B$ for the case $\Gamma = {\mathbb H}$.
The first two columsn describe the quantum symmetry; the column $d_2(B)$
gives the image of the quantum symmetry, and hence the anomaly that
can be resolved; and the last column gives the physical theory
equivalent to $[X/\Gamma]_B$.  (No discrete torsion is possible for
this choice of $\Gamma$.)
}
\begin{tabular}{cc|c|c}
\hline
$B(a)$ & $B(b)$ & $d_2(B)$ & $[X/\Gamma]_B$ \\ \hline
1 & 1 & $-$ & $[X/G] \coprod [X/G]_{\rm d.t.}$ \\
-1 & 1 & $\langle a \rangle, \langle ab \rangle$ & $[X/\langle b \rangle]$ \\
1 & -1 & $\langle b \rangle, \langle a b \rangle$ & $[X/\langle a \rangle]$ \\
-1 & -1 & $\langle a \rangle, \langle b \rangle$ & $[X/\langle ab \rangle]$
\end{tabular}
\end{center}
\end{table}

The first row of table~\ref{table:h} describes the case of no quantum symmetry.
In this case, all of the nontrivial quantum symmetries can be used
to resolve an anomaly, and in each case, the resulting physical
theory $[X/\Gamma]_B$ is equivalent to an orbifold which does not 
intersect an anomalous subgroup.  Again, we see that the Wang-Wen-Witten
prescription works.

For our next resolution of the anomalous $[X/G]$ orbifold,
for the same $G$ as before, 
we take $\Gamma = {\mathbb Z}_2 \times {\mathbb Z}_4$,
\begin{equation}
1 \: \longrightarrow \: {\mathbb Z}_2 \: \longrightarrow \:
{\mathbb Z}_2 \times {\mathbb Z}_4
\: \longrightarrow \: {\mathbb Z}_2 \times {\mathbb Z}_2
\: \longrightarrow \: 1.
\end{equation}
We list all possibilites for the quantum symmetry in table~\ref{table:z2z4}.

\begin{table}[h]
\begin{center}
\caption{\label{table:z2z4}
A list of all possible quantum symmetries, anomalies resolved, and
corresponding orbifolds $[X/\Gamma]_B$ for the case $\Gamma = {\mathbb Z}_2
\times {\mathbb Z}_4$.
The first two columsn describe the quantum symmetry; the column $d_2(B)$
gives the image of the quantum symmetry, and hence the anomaly that
can be resolved; and the last two columns give the physical theory
equivalent to $[X/\Gamma]_B$, for either choice of discrete torsion
in the $\Gamma$ orbifold.
}
\begin{tabular}{cc|c|c|c}
\hline
$B(a)$ & $B(b)$ & $d_2(B)$ & $[X/\Gamma]_B$ w/o d.t. & $[X/\Gamma]_B$
with d.t. \\ \hline
1 & 1 & $-$ & $[X/G] \coprod [X/G]$ & $[X/G]_{\rm d.t.} \coprod [X/G]_{\rm d.t.}$ \\
-1 & 1 & $\langle ab \rangle$ & $[X/\langle b \rangle]$ & 
$[X/\langle b \rangle]$ \\
1 & -1 & $\langle b \rangle, \langle ab \rangle$ & $[X/\langle a \rangle]$ &
$[X/\langle a \rangle]$ \\
-1 & -1 & $\langle b \rangle$ & $[X/\langle ab \rangle]$ & $[X/\langle ab \rangle]$
\end{tabular}
\end{center}
\end{table}

The details here are different, but we see the same overall pattern.
As before, the first row corresponds to the case of no quantum symmetry.
In each of the next three rows, we see that it is possible to resolve
an anomaly, and if one chooses $B$ such that $d_2(B)$ describes the
anomaly, then the resulting physical theory
$[X/\Gamma]_B$ is equivalent to an orbifold by a subgroup which
does not contain the anomalous subgroup.  As before, Wang-Wen-Witten
works.

So far we have picked `minimal' resolutions.
For our next resolution of the anomalous $[X/G]$ orbifold,
for the same $G$ as before, 
we take $\Gamma = {\mathbb Z}_2 \times {\mathbb H}$, 
\begin{equation}
1 \: \longrightarrow \: {\mathbb Z}_2 \times {\mathbb Z}_2 \: \longrightarrow \:
{\mathbb Z}_2 \times {\mathbb H} \: \longrightarrow \:
{\mathbb Z}_2 \times {\mathbb Z}_2 \: \longrightarrow \: 1,
\end{equation}
so that in effect
there is an extra ${\mathbb Z}_2$.
We list all possibilities for the quantum symmetry in
table~\ref{table:z2h}.

\begin{table}[h]
\begin{center}
\caption{\label{table:z2h}
A list of all possible quantum symmetries, anomalies resolved, and
corresponding orbifolds $[X/\Gamma]_B$ for the case $\Gamma = {\mathbb Z}_2 {\mathbb H}$.
The first two columsn describe the quantum symmetry; the column $d_2(B)$
gives the image of the quantum symmetry, and hence the anomaly that
can be resolved; and the last column gives the physical theory
equivalent to $[X/\Gamma]_B$.
}
\begin{tabular}{cc|c|c}
\hline
$B(a)$ & $B(b)$ & $d_2(B)$ & $[X/\Gamma]_B$ \\ \hline
1 & 1 & $-$ & $\coprod_2 \left( [X/G] \coprod [X/G]_{\rm d.t.} \right)$ \\
-1 & 1 & $\langle a \rangle, \langle ab \rangle$ & $\coprod_2 [X/\langle b \rangle]$ \\
1 & -1 & $\langle b \rangle, \langle a b \rangle$ & $\coprod_2 [X/\langle a \rangle]$ \\
-1 & -1 & $\langle a \rangle, \langle b \rangle$ & $\coprod_2 [X/\langle ab \rangle]$
\end{tabular}
\end{center}
\end{table}

The details are different, but the pattern is the same:
by picking $B$ such that the anomaly is described by $d_2(B)$,
the resulting physical theory is non-anomalous, as predicted by
Wang-Wen-Witten.

\section{Conclusions}

In this article we have reviewed work on decomposition,
the observation that sometimes one local quantum field theory
is equivalent to a disjoint union of other local quantum field theories,
known as universes.
This arises when a $d$-dimensional quantum field theory has a $(d-1)$-form
symmetry.  After reviewing examples of decomposition and its properties
(such as multiverse interference effects), we discussed the application to
the anomaly-resolution procedure of Wang-Wen-Witten \cite{Wang:2017loc}.

\begin{acknowledgement}
We would like to thank the many collaborators we have worked with over
the years in making sense of decomposition and its various applications.
Two in particular stand out:  Tony Pantev, for frequent collaborations on
this and related subjects going back roughly twenty years now, and
Simeon Hellerman, for early work on resolving technical challenges
in making sense of strings on stacks that led to this work.
We would also like to specifically thank A.~Cherman, S.~Hellerman, T.~Jacobson,
Y.~Tanizaki, and M.~\"Unsal for discussions of locality and
cluster decomposition and their role in these theories.
We gratefully acknowledge NSF support, most recently via NSF grant
PHY-2014086.
\end{acknowledgement}

\bigskip
%
 \bibliographystyle{spmpsci}
 \bibliography{refs}

\end{document}